# PARABOLIC EQUATIONS ON DIGITAL SPACES. SOLUTIONS ON THE DIGITAL MOEBIUS STRIP AND THE DIGITAL PROJECTIVE PLANE

Alexander V. Evako

Moscow State University of Electronics and Automatics, Dianet,
Volokolamskoe Sh. 1, kv. 157, 125080 Moscow, Russia
Tel/Fax: 095 158 2939, e- mail: evakoa@mail.ru.

**Abstract.**
In this work, we define a parabolic equation on digital spaces and study its properties. The equation can be used in investigation of mechanical, aerodynamic, structural and technological properties of a Moebius strip, which is used as a basic element of a new configuration of an airplane wing. Condition for existence of exact solutions by a matrix method and a method of separation of variables are studied and determined. As examples, numerical solutions on Moebius strip and projective plane are presented.

## 1   Introduction.

The paper presents results of the investigation of partial differential equations on digital spaces published in 1997-1999 in papers [2-5] (in Russian language).
Non-orientable surfaces such as a Moebius strip, Klein bottle and projective plane have recently attracted many scientists from other fields. We mention here, among the others, physics, where a considerable interest has emerged in studying lattice models on non-orientable surfaces as new challenging unsolved lattice-statistical problems and as a realization and testing of predictions of the conformal field theory.
In a joint Russian-French-German project [7] a Moebius strip is proposed as a basic element of an airplane wing. The project consists of three stages. At the first stage, problems of improving aerodynamic characteristics of airplane wing, at the second one, problems of improving the structural and stiffness characteristics of main load-carrying elements and at the third stage, problems of improving efficiency of the industrial and household mixers are solved. The common moment for all these stages is the use of surfaces of one-side topology (Moebius strip-type) in the main structural elements.
Many important technical and physical properties of Moebius-type structural elements can be described by solutions of partial differential equations (PDE), where a Moebius strip serves as a domain.
A Moebius strip and other non-orientable surfaces can be presented in a digital form in the frame of digital topology. This theory has been developed to provide a sound mathematical background for image processing operations [8]. Digital topology plays an important role in analyzing n-dimensional digitized images arising in computer graphics as well as in many areas of science including neuroscience, medical imaging, industrial inspection, geoscience and fluid dynamics. Concepts and results of digital topology are used to specify and justify some important low-level image processing algorithms, including algorithms for thinning, boundary extraction, object counting, and contour filling. So it seems reasonable to define and study partial differential equations on digital spaces.
The material to be presented below begins with the definitions of a digital space and digital n-surface [1, 4-5]. We describe digital spheres, a Moebius strip and a projective plane. The important feature of an n-surface is a similarity of its properties with properties of its continuous counterpart in terms of algebraic topology. For example, the Euler characteristics and the homology groups of digital n-spheres, a Moebius strip and a Klein bottle are the same as ones of their continuous counterparts. Then we define a parabolic differential equation on a digital space and study some its properties. We investigate a stability of a solution and as example, give a numerical solution of a parabolic equation on a Moebius strip and a projective plane.

Since analytic solutions of PDE can be obtained only in simple geometric regions, for practical problems, it is more reasonable to use computational or numerical solutions. We can do this by implementing as domains digital spaces, which are discreet counterparts of continuous spaces and by transferring partial differential equations from a continuous area into discrete one. In the finite difference method for solving partial differential equations in two- and three dimensions, a two- or three-dimensional continuous domain is replaced by a grid. In fact, this grid is a digital model of a continuous space. There arises a serious problem because in most of cases, the grid is not a correct two- or three- dimensional space in terms of digital topology and, therefore cannot properly model the continuous domain. There is a principal difference between PDE on a digital space and PDE on a grid: a digital space is a grid itself and cannot be changed, while a grid in the net method can be chosen in a variety of ways. Distinctions between the differential equations on discrete and continuous spaces are also essential. One of differences is stipulated by the fact that a digital space can have just a few points. Another serious difference is linked to the existence of the natural least length in a digital space, defined by the length of the edge connecting two adjacent points of the space. In application to wave processes it means a lack of indefinitely short waves and indefinitely high frequencies, that is the lack of the factors frequently conducting to divergences.

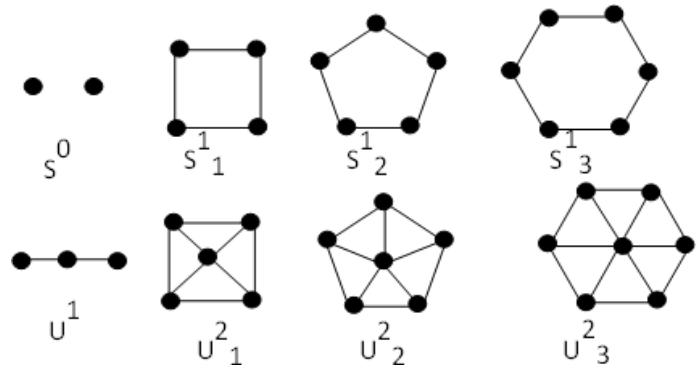

Figure 1. Zero- and one-dimensional spheres $S^0, S^1_1, S^1_2, S^1_3$, and one- and two-dimensional balls $U^1, U^2_1, U^2_2, U^2_3$.

## 2   Digital n-surfaces.

In order to make this paper self-contained, we summarize the necessary information from previous papers [1, 4-5].

A digital space G is a simple undirected graph G=(V,W) where $V=(v_1,v_2,...v_n,...)$ is a finite or countable set of points, and $W = ((v_p v_q),....)$ is a set of edges. Topological properties of G as a digital space in terms of adjacency, connectedness and dimensionality are completely defined by set W.

Let G and v be a digital space and a point of G. The subspace O(v) containing all neighbors of v (without v) is called the rim of point v in G. The subspace U(v) containing O(v) as well as point v is called the ball of point v in G. Apparently, U(v)-

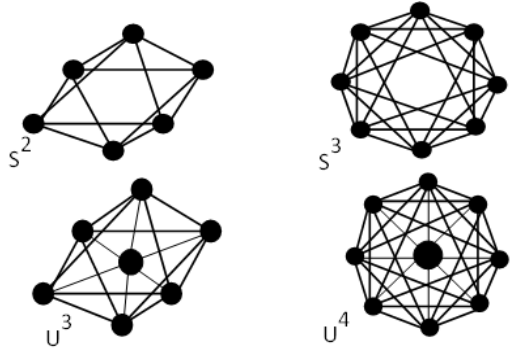

Fig. 2. Two-and three-dimensional spheres $S^2$ and $S^3$ and three- and four-dimensional balls $U^3$ and $U^4$.

v=O(v).

**Definition 2.1.**

The digital 0-dimensional surface $S^0(a,b)$ is a disconnected graph with just two points a and b. For n>0, a digital n-dimensional surface $G^n$ is a nonempty connected graph such that, for each point v of $G^n$, O(v) is a finite digital (n-1)-dimensional surface .

Point v and its ball U(v) in a digital space G are called a normal n-dimensional point and a normal n-dimensional ball respectively if the rim O(v) is an (n-1)-dimensional surface.

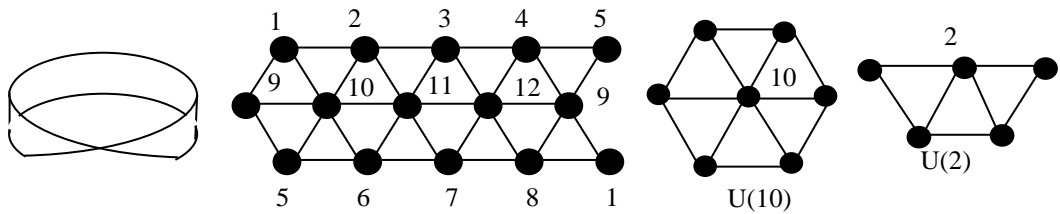

Fig. 3. The Moebius strip is formed by interior points 9, 10, 11, 12 and boundary points 1-8. All interior points are two-dimensional ones. Boundary points form an one-dimensional sphere.

The digital 0-dimensional surface $S^0(a,b)$ is called the digital 0-dimensional sphere. Figure 1 depicts zero- and one-dimensional spheres (circles) $S^0$, $S^1_1$, $S^1_2$, $S^1_3$, and one- and two-dimensional balls $U^1$, $U^2_1$, $U^2_2$, $U^2_3$. Two-and three-dimensional spheres $S^2$ and $S^3$ and three- and four-dimensional balls $U^3$ and $U^4$ are shown in figure 2. A Moebius strip depicted in fig. 3 consists of twelve points. Points 9, 10, 11 and 12 are interior two-dimensional points, points 1-8 are boundary points which form a one-dimensional sphere. The first projective plane depicted in fig. 4 consists of sixteen points, the second one is the minimal digital projective plane with eleven points. It is easy to see that the rim of any point in $P^2_1$ and $P^2_2$ is a circle.

In the finite difference method for solving partial differential equations in two- and three dimensions, a two- or three-dimensional continuous domain is replaced by a grid. In fact, this grid has to be a digital model of a continuous space. However, in most of cases, the grid is not a correct two- or three- dimensional space in terms of digital n-surfaces. For example, consider a standard two-dimensional grid G (fig. 5) often used in finite-difference schemes. As one can see, the neighborhood O(v) of any point v consists of four non-adjacent points and, therefore, is not a one-dimensional sphere. Hence, G is not a part of a digital two-dimensional plane, but rather can be seen as a collection of one-dimensional segments. Grid H is a part of a digital plane because the rim of any point is a digital 2-sphere containing six points. Thus. H is a correct grid which should be used in finite-difference schemes.

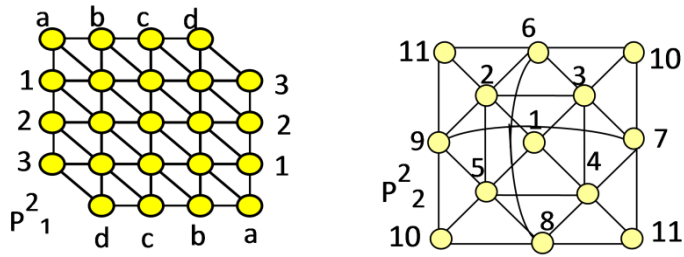

Fig. 4. Two-dimensional projective planes $P^2_1$ and $P^2_2$ with sixteen and eleven points. $P^2_2$ is the minimal projective plane obtained from $P^2_1$ by contractible transformations.

## 3  Properties of a parabolic equation on a digital space.

The PDE on a digital space can be set by analogy with an explicit discretization scheme used for a numerical solution of the partial differential equations. For example, consider the heat equation (1) in 2D. Two-dimensional space grid is shown in fig.6. Using the approximation of space and time derivatives, we obtain the difference equation (2).

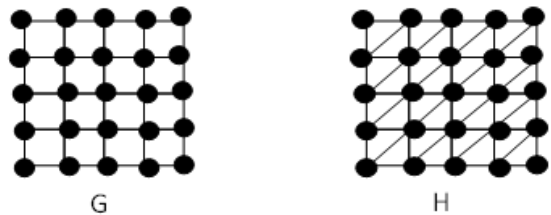

Figure 5. Grid G is not a part of a digital plane. Grid H is a part of a digital plane.

$$\frac{\partial f}{\partial t} = k(\frac{\partial^2 f}{\partial x^2} + \frac{\partial^2 f}{\partial y^2}) + Q, \qquad k > 0 \qquad (1)$$

$$f_{i,j}^{t+1} = f_{i,j}^t + \frac{k\Delta t}{\Delta y^2}(f_{i,j-1}^t - 2f_{i,j}^t + f_{i,j+1}^t) +$$

$$\frac{k\Delta t}{\Delta x^2}(f_{i-1,j}^t - 2f_{i,j}^t + f_{i+1,j}^t) + Q_{i,j}\Delta t \qquad (2)$$

Based on these conditions, we define a differential parabolic equation on digital space G.

**Definition 3.1.**

Let G be a digital space with points $(v_1, v_2, v_3, \ldots v_n)$. A differential parabolic equation on G is the set of n equations of the form

$$f_p^{t+1} = \sum_{v_k \in U(v_p)} c_{pk} f_k^t + q_p^t, \quad p=1,\ldots n, \quad \forall c_{pk} \geq 0, \quad \sum_{v_p \in U(v_k)} c_{pk} = 1, \quad k=1,\ldots n \qquad (3)$$

Here $f_k^t$ and $q_p^t$ are values of the functions in points $v_k$ and $v_p$ of G at a moment t, $f_p^{t+1}$ is the value of the function in point $v_p$ at moment t+1, $c_{pk}$ are coefficients, the summation is produced over all points $v_k$ belonging to the ball $U(v_p)$ of point $v_p$. If points $v_p$ and $v_k$ are non-adjacent [2, 3] then $c_{pk}=0$. Then set (3) can be written in the form

$$f_p^{t+1} = \sum_{k=1}^n c_{pk} f_k^t + q_p^t, \quad p=1,\ldots n, \quad \forall c_{pk} \geq 0, \quad \sum_{p=1}^n c_{pk} = 1, \quad k=1,\ldots n \qquad (4)$$

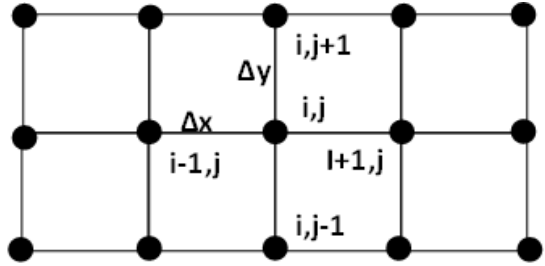

Figure 6. Two-dimensional space grid.

If $q_k^t = 0$ for any k and t then

$$f_p^{t+1} = \sum_{k=1}^n c_{pk} f_k^t, \quad p=1,\ldots n, \quad \forall c_{pk} \geq 0, \quad \sum_{p=1}^n c_{pk} = 1, \quad k=1,\ldots n \qquad (5)$$

is called a homogeneous differential parabolic equation on G.

In general, coefficients $c_{pk}$ and function $q_k^t$ depend on p, k and t. Notice that equation (1) does not depend explicitly on the dimension of G and can be applied to a digital space of any dimension. All dimensional features are contained in digital space G.

Equations (4-5) can be presented in the matrix form

$$f^{t+1} = Cf^t + g^t, \quad f^{t+1} = Cf^t$$

$$f^{t+1} = \begin{bmatrix} f_1^{t+1} \\ f_2^{t+1} \\ * \\ f_n^{t+1} \end{bmatrix}, \quad C = \begin{bmatrix} c_{11} & c_{12} & * & c_{1n} \\ c_{21} & c_{21} & * & * \\ * & * & * & * \\ c_{n1} & * & * & c_{nn} \end{bmatrix}, \quad f^t = \begin{bmatrix} f_1^t \\ f_2^t \\ * \\ f_n^t \end{bmatrix}, \quad g^t = \begin{bmatrix} g_1^t \\ g_2^t \\ * \\ g_n^t \end{bmatrix}, \quad (6)$$

Equation (5) along with initial values $f^0_p$, p=1,2,…n, is called the initial value problem for the parabolic differential equation on a digital space G. Define the stability of equation (5) using a standard approach. For stability, we will need a norm. Hence, for $f^t$, the norm is defined as $\|f^t\| = |f_1^t| + \ldots |f_n^t|$. Equation (5) is called stable according to initial values if there exists a positive M such that $\|f^t\| \leq M \|f^0\|$ for all t. It is easy to see that this equation is stable.

**Proposition 3.1.**
Equation (5) is stable.
**Proof.**

Consider $\|f^{t+1}\|$. Then

$$\|f^{t+1}\| = \sum_{p=1}^n |f_p^{t+1}| = \sum_{p=1}^n |\sum_{k=1}^n c_{pk} f_k^t| \leq \sum_{p=1}^n \sum_{k=1}^n c_{pk}|f_k^t| = \sum_{p=1}^n \sum_{k=1}^n c_{pk}|f_k^t| = \sum_{k=1}^n \sum_{p=1}^n c_{pk}|f_k^t| = \sum_{k=1}^n |f_k^t| = \|f^t\|.$$

Since $\|f^{t+1}\| \leq \|f^t\|$ then $\|f^{t+1}\| \leq \|f^0\|$. It completes the proof. □

In equation (5), consider the sum $S^t = \sum_{p=1}^n f_p^t$ of values of the function $f^t$ on all points of the digital space G.

**Proposition 3.2.**
In equation (5) sums of values of the function $f^t$ on all points of the space G do not depend on t.
**Proof.**

$$S^{t+1} = \sum_{p=1}^n f_p^{t+1} = \sum_{p=1}^n \sum_{k=1}^n c_{pk} f_k^t = \sum_{p=1}^n \sum_{k=1}^n c_{pk} f_k^t = \sum_{k=1}^n \sum_{p=1}^n c_{pk} f_k^t = \sum_{k=1}^n f_k^t$$

It completes the proof. □

At rather large times t, the form of a solution will be determined by a limit form of a matrix $C^t$. It is of some interest to clarify a behavior of matrix $C^t$ as t→∞. Call this limit as a final matrix $C^\infty$. The final matrix $C^\infty$ converts initial values $f^0_k$ into final values $f^\infty_k$.

In the equation (6), the square (n x n)-matrix $C=\{c_{pk}\}$ of coefficients is a stochastic matrix (or, more correct, a transposed stochastic matrix) which properties are well known [6]. Such matrices are used in Markov processes. Let's consider some applications of such approach for a solution of the initial value problem.

Remind that square matrix C is called indecomposable if it can not be converted into the form

$$C = \begin{bmatrix} C_{11} & 0 \\ C_{12} & C_{22} \end{bmatrix},$$

by simultaneous relabeling of rows and columns. Here $C_{11}$ and $C_{22}$ — square matrices. Square matrix C is called primitive if it can not be turned into a cyclic form

$$C = \begin{bmatrix} 0 & 0 & \ldots & 0 & C_{1h} \\ C_{21} & 0 & \ldots & 0 & \\ 0 & C_{32} & \ldots & 0 & \\ \ldots & \ldots & \ldots & \ldots & \ldots \\ & \ldots & & C_{h,h-1} & 0 \end{bmatrix},$$

by simultaneous relabeling of rows and columns. Here $C_{kp}$ are square matrices.

**Proposition 3.3..**

If matrix C in (6) is indecomposable and primitive, then the final solution $f^\infty$ of equation (6) as $t \to \infty$ at any initial values is stationary, not time-dependent (in each point of space, a value of function f is constant, does not depend on time t), is determined only by the sum of values of function f in all points of the space G and has the form:

$$f^\infty = C^\infty f^0, \quad \text{where}$$

$$f^\infty = S \bullet \begin{bmatrix} c_1 \\ c_2 \\ * \\ c_n \end{bmatrix}, \quad C^\infty = \begin{bmatrix} c_1 & c_1 & * & c_1 \\ c_2 & c_2 & * & c_2 \\ * & * & * & * \\ c_n & c_n & * & *c_n \end{bmatrix}, \quad f^0 = \begin{bmatrix} f_1^0 \\ f_2^0 \\ \ldots \\ f_n^0 \end{bmatrix}, \quad \sum_{k=1}^{n} f_k^0 = S,$$

$$\sum_{k=1}^{n} c_k = 1, \quad \forall c_k > 0.$$

**Proof.**

Since matrix C is indecomposable and primitive, then it has a simple maximum eigenvalue 1, and there are no other complex eigenvalues which modulus are equal 1 [6]. As it was shown in [4] matrix C converges to a limit stochastic matrix $C^\infty$ as $t \to \infty$ which can be presented in the following form:

$$C^\infty = \begin{bmatrix} c_1 & c_1 & * & c_1 \\ c_2 & c_2 & * & c_2 \\ * & * & * & * \\ c_n & c_n & * & *c_n \end{bmatrix}, \quad \sum_{k=1}^{n} c_k = 1, \quad \forall c_k > 0.$$

In this matrix, all elements are strictly positive, the sum of all elements in each column is equal to 1. Besides, the column is the eigenvector of matrix C appropriate to the eigenvalue 1. Since $\sum_{\kappa=1}^{n} f_k^{t+1} = \sum_{\kappa=1}^{n} f_k^t = S$, then the equation (5-6) as $t \to \infty$ has the solution of the form

$$f^\infty = C^\infty f^0, \quad \text{where}$$

$$f^\infty = S \bullet \begin{bmatrix} c_1 \\ c_2 \\ * \\ c_n \end{bmatrix}, \quad C^\infty = \begin{bmatrix} c_1 & c_1 & * & c_1 \\ c_2 & c_2 & * & c_2 \\ * & * & * & * \\ c_n & c_n & * & *c_n \end{bmatrix}, \quad f^0 = \begin{bmatrix} f_1^0 \\ f_2^0 \\ ... \\ f_n^0 \end{bmatrix}, \quad \sum_{k=1}^{n} f_k^0 = S,$$

$$\sum_{k=1}^{n} c_k = 1, \quad \forall c_k > 0.$$

The proof is complete. □

From these expressions, it is clear that such a solution depends only on S, but not on concrete distribution of values of function f on points of space G in the initial moment t = 0. Besides, any function of the form

$$f = d \begin{bmatrix} c_1 \\ c_2 \\ ... \\ c_n \end{bmatrix}$$

is an eigenvector of C, and a stationary solution of the equation (5-6) because $Cf = CC^\infty f = C^\infty f = f$. That is

$$f_p = \sum_{v_k \in G} c_{pk} f_k.$$

This equation can be considered as an analog of an elliptical equation on a digital space.

Now, obtain a solution of a parabolic equation (6) by the method of a separation of variables. Assume that matrix $C = \{c_{pk}\}$ in (6) can be reduced to a diagonal form by homothetic transformations. As it is known, it is possible to do if and only if C commutes with C', that is CC'=C'C, where C' is transposed in relation to C. In particular, it is possible if C is symmetric. In this case, there are n eigenvalues of the matrix C (some of them can coincide). Besides, the total number of linearly independent eigenvectors $X_1, X_2, ... X_n$ of the matrix C is equal to n.

Let's take the solution f of the equation (6) as a product of two functions, one of which T(t) depends only on time and another $X = \{X(k)\} = \{X(v_k)\}$ only on points of the space. Assume that X is an eigenvector of C and λ is a real eigenvalue of a matrix C corresponding to X.

$$f_p^t = T(t)X(p), \quad f = T(t)X.$$

Substituting this expression in (6), we receive

$$T(t) = \lambda^t, \quad f^t = d\lambda^t X, \quad \text{or} \quad f_p^t = d\lambda^t X(p)$$

Here d is any real number. Taking into account, that the total number of linearly independent eigenvectors $X_k = \{X_k(p)\}$, k=1,2,…n, of C is equal to n, the general solution of the equation (5-6) has the form

$$f^t = \sum_{s=1}^{n} d_s \lambda_s^t X_s, \quad or \quad f_p^t = \sum_{s=1}^{n} d_s \lambda_s^t X_s(p)$$

Here, $d_s$ are any constants, and some eigenvalues can coincide.
We formulate the above consideration in the form of the theorem.

**Proposition 3.4.**

If matrix C in (6) commutes with its transposed matrix C', then the general solution of the equation (6) has the form

$$f^t = \sum_{s=1}^{n} d_s \lambda_s^t X_s, \quad or \quad f_p^t = \sum_{s=1}^{n} d_s \lambda_s^t X_s(p),$$

where $X_k = \{X_k(p)\}$, k=1,2,…n, are independent eigenvectors, $\lambda_k$ are corresponding eigenvalues of a matrix C, $d_k$ are real numbers.

For a particular solution of (5-6), coefficients $d_s$ can be found by using given initial values of the function $f^0 = \{f_p^0\}$ at the initial moment t = 0.

If matrix C commutes with C', indecomposable and primitive, then matrix C has the only positive eigenvalue $\lambda_1 = 1$ of order (algebraic) 1 [6]. Moduli of all other eigenvalues are less 1. It is easy to check that for any eigenvalue $\lambda_k$, $|\lambda_k| \neq 1$, the eigenvector $X_k = \{X_k(p)\}$ corresponding to $\lambda_k$ satisfies the condition

$$\sum_{p=1}^{n} X_k(p) = 0.$$

In this case the solution of equation (5-6) is defined by the expression

$$f_p^t = d_1 X_1(p) + d_2 \lambda_2^t X_2(p) + \ldots + d_n \lambda_n^t X_n(p).$$

Since $\lambda_k^t \to 0$ as $t \to \infty$, k=2,3,…n, then the final solution has the form

$$f_p^\infty = d_1 X_1(p).$$

## 4 A numerical solution of a parabolic equation on a Moebius strip and projective plane

In this section, we consider the parabolic equation on two digital two-dimensional surfaces: a Moebius strip and a projective plane. For simplicity, we will use surfaces with a small number of points.

The Moebius strip depicted in fig. 4 consists of twelve points. Points 9, 10, 11 and 12 are interior two-dimensional points, points 1-8 are boundary points which form a one-dimensional sphere (circle).

$$f_p^{t+1} = \sum_{k=1}^{n} c_{pk} f_k^t, \quad \forall c_{pk} \geq 0, \quad \sum_{p=1}^{n} c_{pk} = 1$$

For interior point p=10, $c_{pp}$=0.82, $c_{ps}$=0.03, s=1, 2, 11, 6, 5, 9. For all other interior points p=9, 11, 12, the structure of coefficients is the same, $c_{pp}$=0.82, $c_{ps}$=0.03, $s\in O(p)$. For boundary point p=2, $c_{pp}$=0.88, $c_{ps}$=0.03, s=1, 10, 11, 3. For all other boundary points p=1, 3-8, the structure of coefficients is similar, $c_{pp}$=0.88, $c_{ps}$=0.03, $s\in O(p)$. Initial values are given as $f^0_1$=12, $f^0_p$=0, p=2,…12. The solutions in boundary point 3 and internal point 10 are presented in fig. 7(a), where t=0,1,…100.

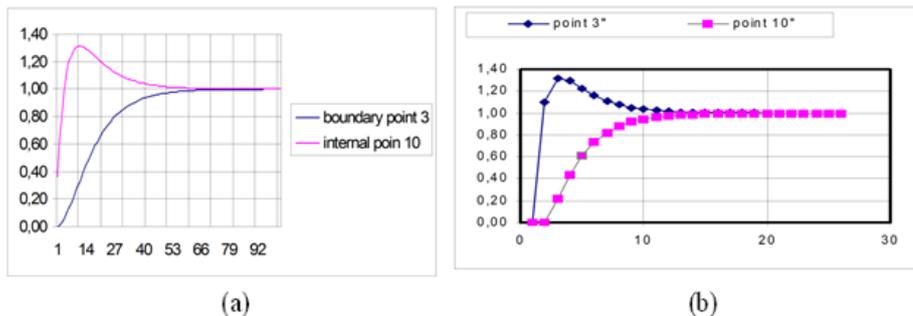

Figure 7. (a) The solution profiles of the parabolic equation: (a) on the Moebius strip in point 3 and point 10. (b) on the projective plane in point 3 and point 10.

In the projective plane $P^2_2$ (fig. 4), all points are interior. Choose $c_{11}$=0.6, $c_{22}$=$c_{33}$=$c_{44}$=$c_{55}$=$c_{1010}$=$c_{1111}$=0.4, $c_{66}$=$c_{77}$=$c_{88}$=$c_{99}$=0.5, all other coefficients $c_{ps}$=0.1, where $v_p\in O(v_s)$. Initial values are given as $f^0_1$=11, $f^0_p$=0, p=2,…10. The solutions in points 3 and 10 are presented in fig. 7(b), where t=0,1,…30.